\documentclass[twocolumn,prd,floatfix,nofootinbib,longbibliography,notitlepage,superscriptaddress]{revtex4-2}

\pdfoutput=1

\usepackage{amsmath, amssymb, amsfonts, amsthm, latexsym, mathrsfs, bbm, slashed}

\usepackage[usenames,dvipsnames]{xcolor}
\usepackage[title,titletoc]{appendix}
\usepackage{graphicx}
  
\usepackage{xifthen}
\usepackage{tensor}
\usepackage{dsfont}
\usepackage{subcaption}
\usepackage[inline]{enumitem}

\usepackage{setspace}
\usepackage[marginal, multiple]{footmisc}
 
\usepackage[T1]{fontenc}
\usepackage[utf8]{inputenc}
\usepackage{lmodern}
\usepackage[normalem]{ulem}

\usepackage[unicode, colorlinks, allcolors=blue!70!black, linktocpage, pdfusetitle]{hyperref}
\definecolor{CiteColor}{rgb}{0.55,0,0}
\hypersetup{citecolor=CiteColor}
\definecolor{RefColor}{rgb}{0,0.5,0}
\hypersetup{linkcolor=RefColor}
\usepackage[all]{hypcap}

\usepackage{orcidlink}

\usepackage{verbatim}

\usepackage{cancel}

\usepackage{microtype}

\usepackage{standalone}
\usepackage{tikz}
\usetikzlibrary{arrows.meta}
\usetikzlibrary{decorations.markings}
\ifdefined\myext
\usetikzlibrary{external}
\fi

\newcommand{\nn}{\nonumber}

\newcommand{\pd}{\partial}

\DeclareMathOperator{\sn}{sn}

\newcommand{\eff}{\text{eff}}

\newcommand{\Eff}{ {S}_{\eff}\cdot {L}}
\newcommand{\Ef}{ {s}_{\eff}\cdot {l}}

\newcommand{\pb}[1]{\left\lbrace  #1   \right\rbrace  }

\newcommand{\figA}{%
\begin{figure}[t]
  \includegraphics[width=8cm]{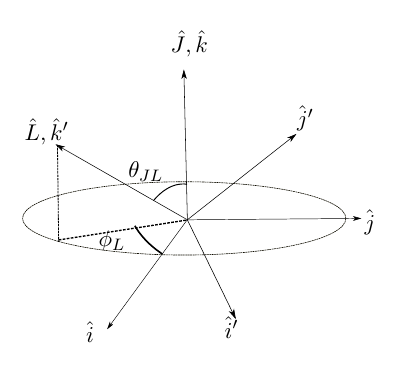}
  \caption{The non-inertial $(i'j'k')$ frame  (centered around
   $\hat{L} \equiv \vec{L}/L$) is displayed along with the inertial $(ijk)$ frame (centered around $\hat{J} \equiv \vec{J}/J$).} 
  \label{fig:1} 
\end{figure}
}

\newcommand{\figB}{%
\begin{figure*}
  \includegraphics[width=12cm]{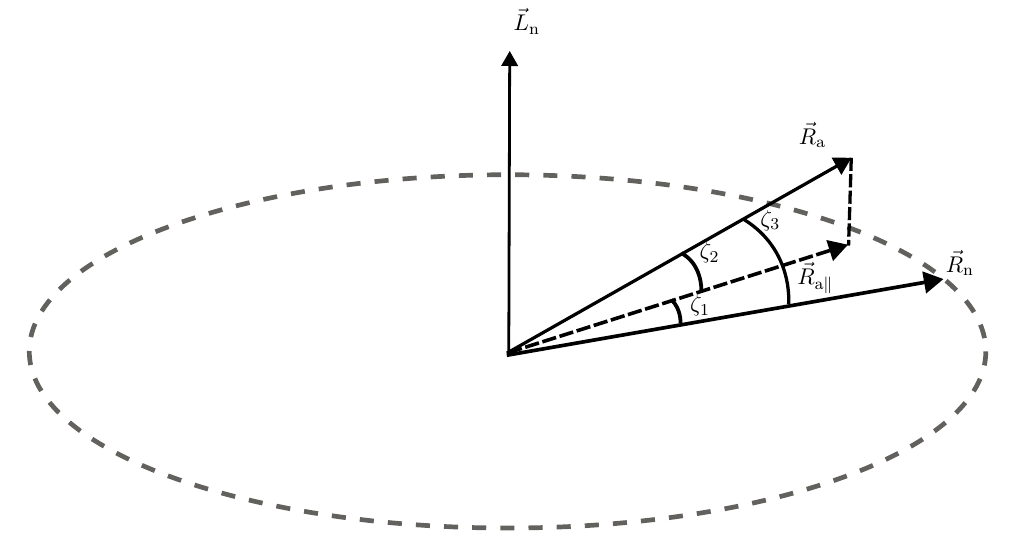}
  \caption{Difference between the numerical and analytical position vector $\vec{R}$ being split in two 
  perpendicular components.} 
  \label{fig:2} 
\end{figure*}
}

\begin{document}

\title{Closed-form solutions of spinning, eccentric binary
black holes at 1.5 post-Newtonian order}

\newcommand{\UMiss}{\affiliation{Department of Physics and Astronomy, The University of Mississippi, University, MS 38677, USA}}

\newcommand{\LUX}{\affiliation{LUX, Observatoire de Paris, Université PSL, Sorbonne Université, CNRS, 92190 Meudon, France}}

\author{Rickmoy Samanta\,\orcidlink{0000-0002-4735-7774}}
\affiliation{Indian Statistical Institute, 203 B. T. Road, Kolkata 700108, India}
\affiliation{ Birla Institute of Technology and Science Pilani, Hyderabad 500078, India}
\author{Sashwat Tanay\,\orcidlink{0000-0002-2964-7102}}
\email{sashwat.tanay@obspm.fr}
\LUX
\UMiss
\author{Leo C.~Stein\,\orcidlink{0000-0001-7559-9597}}
\UMiss

\hypersetup{pdfauthor={Tanay, Cho, and Stein}}

\begin{abstract}
The closed-form solution of the 1.5 post-Newtonian (PN)
 accurate binary black hole (BBH) Hamiltonian system
has proven to be evasive for a long time since the introduction of the 
system in 1966. 
Solutions of the PN BBH systems with arbitrary parameters (masses, spins,
eccentricity) are required for modeling the gravitational waves (GWs) emitted by them. 
Accurate models of GWs are crucial for their detection by LIGO/Virgo and LISA.
Only recently, two solution methods for solving the
 BBH dynamics were proposed in 
Ref.~[\prd~100, 044046 (2019)] 
(without using action-angle variables), and Refs.~[\prd~103, 064066 (2021),
\prd~107, 103040 (2023)] (action-angle based).
 This paper combines the ideas 
laid out in the above articles, 
fills the missing gaps and compiles the two solutions
which are fully 1.5PN accurate. 
We also present a public \textsc{Mathematica} package
\texttt{BBHpnToolkit} which implements these
two solutions and compares them with the result of numerical
integration of the evolution equations.
The level of agreement between these solutions
  provides a  numerical verification
for all the five action variables constructed
 in Refs.~[\prd~103, 064066 (2021),
\prd~107, 103040 (2023)]. This paper hence serves as 
a stepping stone for pushing the action-angle-based solution to 
2PN order via canonical perturbation theory. 
\end{abstract}

\maketitle

\section{Introduction}
Construction of accurate gravitational wave (GW) 
templates (or models) has been crucial to the 
the  GW detections that have taken place so far since 2015
\cite{ LIGOScientific:2018mvr, Abbott:2020niy, LIGOScientific:2017vwq}. This is so because the method of 
matched filtering for GW detection requires as one of the inputs, the theoretical templates
of the GW signal to be detected.
Post-Newtonian (PN) theory serves as a useful framework within which GWs from binary black holes (BBHs) are modeled
when the system is in its initial and longest-lived inspiral stage \cite{Blanchet:2013haa}.
 At this stage, the two black holes (BHs) of the BBH
are far apart and move slowly around a common center. This is also referred to as the PN regime.
In the PN framework, quantities of interest are presented in a PN power series in the small PN paramter
(ratio of the typical speed of the system and that of light).
As is typical of power series, higher-order corrections are of smaller magnitudes and carry higher PN orders.
Since GWs from a BBH are functions of the positions-momenta of the source, modeling the 
positions-momenta of the BBH system is crucial for constructing the GW templates.
This paper deals with the 
construction of 1.5PN accurate closed-form solutions of the BBH system.

Since we restrict ourselves to 1.5PN order, the dissipative effects on the BBH dynamics
due to GW emission don't enter the picture; they show up at 2.5PN.
The conservative dynamics of the  system can be described with the PN Hamiltonian framework,
wherein the system possesses a Hamiltonian that is a
 function of the system's canonical coordinates \cite{Schafer:2018kuf}.
The leading PN order effect is simply that of two point masses  moving under mutual Newtonian 
gravitational attraction whose Hamiltonian treatment is a textbook subject matter.
Such systems, move on a closed ellipse if they are bound.
The next level of sophistication is at 1PN order wherein 1PN effects are added to the 
above Newtonian system. At this level, spin effects are ignored (they enter at 1.5PN).
Ref.~\cite{1985AIHS...43..107D} provided the quasi-Keplerian parametric solution for this system;
the trajectory is no longer a closed ellipse and 
the system features the advance of periastron. The 
orbit still remains confined in a plane due 
to the constancy of the angular momentum vector.

Moving further up the PN ladder, we encounter the 1.5PN system whose
Hamiltonian was proposed in Ref.~\cite{Barker:1966zz}. At 
1.5PN order, spin effects come into play for the first time
via a spin-orbit interaction  (linear-in-spin), while the spin-spin interaction 
terms enter at 2PN. Via numerical integration of the resulting 
equations of motion (EOMs), it is seen that the orbital plane precesses; the orbital
angular momentum is constant only in magnitude, but not in direction. This system displays the rich interplay 
of non-zero BH spins, periastron precession, along with spin and orbital-plane precession. We concern 
ourselves with this system in this paper.

Over the past decades, solutions to the 
dynamics of the spinning BBH system (at 1.5PN order or higher)
have been constructed by many groups 
\cite{tiwari_gopu,2015PhRvL.114h1103K,
2015PhRvD..92f4016G,2018PhRvD..98d4015H,2017PhRvD..95j4004C,2010PhRvD..81l4001K,2018PhRvD..98j4043K,
Konigsdorffer:2006zt, Konigsdorffer:2005sc},
but most of them worked under some simplifying specialization like
only one BH spinning, equal masses, small eccentricity, orbit-averaging, etc.
Two recent breakthroughs have occurred on the front of finding solutions
to the \textit{most general} 1.5PN BBH system without any simplifying assumptions
where the qualifier ``most general'' indicates a system with arbitrary values of
masses, spins, and eccentricity, while 
still falling under the PN regime\footnote{In the PN approximation,
the spin magnitude of a maximally spinning BH 
is 0.5PN order smaller than the BH's orbital
angular momentum.}.
The first breakthrough was made by Cho and Lee in Ref.~\cite{cho2019brd} where they succeeded in
integrating the EOMs of the system; the 1PN term in the Hamiltonian was ignored throughout 
for simplicity.
The second breakthrough came in the form of Refs.~\cite{Tanay:2020gfb, Tanay:2021bff}, where the authors
evaluated all the action variables (actions as in action-angle (AA) variables) and laid out 
a scheme on how to construct the AA-based solution of the 
system\footnote{See Ref.~\cite{Tanay:2022rlz} for a pedagogical
introduction to the action-angle method of solution.}. 
Against this backdrop, this paper aims to target the
following objectives
\begin{enumerate}
\item Re-present the solution of Ref.~\cite{cho2019brd} but with the 1PN terms included. We will call it the Standard Solution.
\item Present a systematic procedure to construct the AA-based solution. 
As we will see later, the construction of the AA-based 
solution requires the Standard Solution as one of the inputs.

\item Make available \texttt{BBHpnToolkit},
 a public \textsc{Mathematica} package \cite{MMA1} which 
 (1) implements the Standard Solution, the AA-based solution, and 
the numerical solution  
(2) gives the numerical 
values of all five frequencies 
(rate of increase of the angle variables) of a given BBH system, 
(3) computes the Poisson brackets (PBs)
 between any two functions of the phase-space variables.
\end{enumerate}
Let us mention that the AA-based solution 
provides a significant advantage over the Standard solution.
This is so because
while extending the Standard Solution to 2PN appears quite difficult, 
the AA-based solution should be extendable to 2PN
via canonical perturbation theory.
  This paper  assembles the ideas put forth in 
Refs.~\cite{Cho:2019brd, Tanay:2020gfb, Tanay:2021bff}, and provides a solid platform 
from where one can springboard to push the 1.5PN solutions to 2PN order using canonical perturbation
theory.

The organization of the paper is as follows.
In Sec.~\ref{setup}, we introduce the phase space, the Hamiltonian, 
the EOMs of the system as well as the concept of AA variables.
In Sec.~\ref{hmsol}, we re-present the solution of Ref.~\cite{Cho:2019brd} with the
1PN terms included (Standard Solution). In Sec.~\ref{aasol}, 
we lay out an in a step-by-step
fashion, the strategy to construct the alternative AA-based solution.
In Sec.~\ref{cmp}, we introduce the \textsc{Mathematica} package \texttt{BBHpnToolkit}
that we release with this paper. We also make comparisons of our analytical solutions 
with the numerical one, before summarizing in Sec.~\ref{summary}.

\section{The setup}  
 \label{setup}

This paper is the culmination of the research initiated in Refs.~\cite{cho2019brd},
\cite{Tanay:2020gfb} and \cite{Tanay:2021bff}. Since the notations and conventions used in the
first article differs from those in the other two articles, the notations and conventions
of this paper are a mix of the two types.

The system in consideration is  a BBH system in the PN approximation. It consists of 
two BHs of masses $m_1$ and $m_2$ such that the relative separation vector 
of the first BH from the second one 
is $\vec{R}$. We choose to work in the center-of-mass frame 
throughout wherein the momenta of the first BH is equal to the negative of the other and is
represented by $\vec{P}$. The total angular momenta of the BBH system is 
$\vec{J}  = \vec{L} + \vec{S}_1 + \vec{S}_2$, where $\vec{L} \equiv \vec{R} \times \vec{P}$,
and $\vec{S}_1$ and $\vec{S}_2$ are the spin angular momenta of the two BHs.
We also define the 
total mass $ M = m_1+m_2$,
the reduced mass  $\mu \equiv m_{1} m_{2} / M$,  
the symmetric mass ratio $\nu \equiv \mu / M$, and
  $\sigma_{1} \equiv 1+3 m_{2} / 4 m_{1}$ and $\sigma_{2} \equiv 1+3 m_{1} / 4 m_{2}$,
  along with 
 \begin{align}
  \vec{S}_{\mathrm{eff}} \equiv \sigma_{1} \vec{S}_{1}+\sigma_{2} \vec{S}_{2}  .   \label{define_Seff}
  \end{align}
As usual, $G$ and $c$ denote the gravitational constant and the speed of light, respectively.
Finally, note that  we represent the physical time with $t'$, whereas
$t \equiv t'/(GM)$ is reserved for the scaled time. Dots represent a derivative 
 taken with respect to $t$. The norm of any 3D vector will be denoted by the same letter 
 as the vector but without the arrow, i.e. $V \equiv |\vec{V}|  $. 
 The unit vector $\vec{V}/V$ corresponding to a vector $\vec{V}$ is denoted by the use of
 a hat $\hat{V} \equiv \vec{V}/V $.

We now present the 1.5PN Hamiltonian of the system using 
scaled variables $\vec{r} \equiv \vec{R}/(GM)$, $\vec{p} \equiv \vec{P}/\mu$,
\begin{align}
H=H_{\mathrm{N}}+H_{1 \mathrm{PN}}+H_{1.5 \mathrm{PN}} + \mathcal{O}\left(c^{-4}\right), 
\end{align}
where the various PN components are
\begin{align}
H_{\mathrm{N}}=& \mu\left(\frac{p^{2}}{2}-\frac{1}{r}\right), \\
H_{1 \mathrm{PN}}=& \frac{\mu}{c^{2}}\left\{\frac{1}{8}(3 \nu-1) p^{4}+\frac{1}{2 r^{2}}\right.\\
&\left.-\frac{1}{2 r}\left[(3+\nu) p^{2}+\nu(\hat{r} \cdot \vec{p})^{2}\right]\right\}, \\
H_{1.5 \mathrm{PN}}=& \frac{2 G}{c^{2} R^{3}} \vec{S}_{\mathrm{eff}} \cdot \vec{L} .    \label{1.5PN-H}
\end{align}
The evolution of any function $g$ is given by its Poisson bracket (PB) with $H$
\begin{align}
\frac{dg}{dt'}  =  \pb{ g, H}.    \label{EOM}
\end{align}

A few basic rules are needed to evaluate the PB between any two functions of the phase space variables
$\vec{R}, \vec{P}, \vec{S}_1$ and $\vec{S}_2$, the first one being the PB
between the phase space variables. The only non-vanishing PBs between the phase-space variables are given by
\begin{align}
\left\{R^i, P_j\right\}=\delta_j^i, \quad \text { and } \quad\left\{S_a^i, S_b^j\right\}=\delta_{a b} \epsilon_k^{i j} S_a^k ,   \label{fundamental PBs}
\end{align}
and those related by the property of anti-symmetry of the PBs
\begin{align}
\{f, g\}=-\{g, f\}  .    \label{anti-symm}
\end{align}
$a, b$ have been used to label the BHs, whereas $i, j, k$ label the components of the
vectors. $\epsilon_k^{i j}$ is the usual Levi-Civita symbol in three dimensions.
This is despite the appearance of lower index $k$, whose sole purpose
is to implement Einstein summation. 
The second rule is the so-called chain rule (with $v_l$'s standing for the phase-space variables)
\begin{align}
\left\{f, g\left(v_l\right)\right\}=\left\{f, v_l\right\} \frac{\partial g}{\partial v_l} .    \label{chain}
\end{align}
Finally, with the aid of Eqs.~\eqref{fundamental PBs},
 \eqref{anti-symm}, and \eqref{chain}, we can
evaluate the PB between any two quantities.
This also lets us evaluate the time derivative of 
any quantity via Eq.~\eqref{EOM}.

Two quantities $f$ and $g$ are said to commute or be in
involution if $\pb{f,g} = 0$.
A system with Hamiltonian $H$ in $2n$
canonical momenta-positions coordinates
$(\vec{\mathcal{P}},\vec{\mathcal{Q}})$  is said to be integrable if
there exists a canonical transformation to new momenta-positions coordinates
$(\vec{\mathcal{J}}, \vec{\theta})$ such that $H$ is a function of
only the $\mathcal{J}$'s, and that all the $\vec{\mathcal{P}}$ and
$\vec{\mathcal{Q}}$ are $2\pi$-periodic functions of the
angle variables $\vec{\theta}$ \cite{fasano}. An integrable system 
possesses $n$ mutually commuting constants, including the Hamiltonian
(Liouville-Arnold theorem).
Another equivalent but practically more useful definition of actions
is given via~\cite{arnold, fasano}
\begin{align}
  \mathcal{J}_{k}=\frac{1}{2 \pi} \oint_{\mathcal{C}_{k}} \Theta
  =\frac{1}{2 \pi} \oint_{\mathcal{C}_{k}}  \mathcal{P}_{i} d \mathcal{Q}^{i}
  \,,
  \label{eq:action_defined}
\end{align}
where $\mathcal{C}_{k}$ is any loop in the $k$th homotopy class on the
$n$-torus defined by the constant values of the $n$ commuting
constants.

Before ending this section, we will touch upon the 
concept of flows under a function $f$ of the 
phase-space variables
and the associated vector fields.
$f( \vec{\mathcal{P}}, \vec{\mathcal{Q}})$ defines a vector field as
$\pb{\cdot, f }=\pd/\pd \lambda$, and its action on another function
$g( \vec{\mathcal{P}}, \vec{\mathcal{Q}})$ is
$\pd g/\pd \lambda  = \pb{g, f }$~\cite{jose}. 
The integral curves of
the vector field is referred to as the flow of the field.
With this, it is easy to see that the equation governing the real-time evolution
of a Hamiltonian system $dg/dt' = \pb{g, H}$ is basically the flow equation
under the Hamiltonian where the flow parameter $\lambda$ is to be taken
as time $t'$. The next two sections deal with two equivalent 
ways of constructing the solution of 1.5PN BBH system.\\

\section{The standard solution}                 
\label{hmsol}

The Standard Solution is contructed 
by integrating the flow equations under the  1.5PN Hamiltonian.
Ref.~\cite{cho2019brd} constructed the solution of the 
1.5PN BBH system but omitted the
 1PN term of the Hamiltonian
entirely in their analysis as it was deemed by the authors to be
simple to deal with. Here we include the 1PN terms too.
Most of the derivation in Ref.~\cite{cho2019brd} goes through,
 even after including the 1PN terms.
  So, we closely follow their approach and do
   not present the entire derivation.
We provide derivation details only for those 
segments where the derivation presented in Ref.~\cite{cho2019brd} 
has to be significantly modified
due to the inclusion of the 1PN terms.
Needless to say, we refer the reader to Ref.~\cite{cho2019brd}
for the derivation of many of the results that we present in this section.
Ref.~\cite{cho2019brd} also has some typos which we fix in this paper.

\subsection{Angles between the $\vec{L}$, $\vec{S}_1$ and $\vec{S}_2$}      \label{mutual_ang_sec}

We introduce scaled variables via $\vec{l} \equiv \vec{L}/(\mu G M)$,
$h \equiv H/\mu$, and $\vec{s}_a \equiv \vec{S}_a/(\mu G M)$, where $a = 1, 2$
serve to distinguish the two BHs.
Let $\kappa_1$, $\kappa_2$ and $\gamma$ denote the angle between the vector pairs
 $(\vec{L}, \vec{S}_1)$, $(\vec{L}, \vec{S}_2)$, and $(\vec{S}_1, \vec{S}_2)$. 
Without loss of generality, we assume that $m_1 > m_2$.
Under the 1.5PN Hamiltonian, $L$, $S_1$, and $S_2$ 
remain constant in time as their PB with the Hamiltonian vanishes.

The solution for $\cos \kappa_1$ and its method of derivation is the same
as that presented in Sec.~III-A of Ref.~\cite{cho2019brd} which was done without
including the 1PN term in the Hamiltonian. Here we directly present that solution.
It reads
\begin{align}
\cos \kappa_{1}(t) &=x_{1}+\left(x_{2}-x_{1}\right)  {\sn}^{2}(\Upsilon(t), \beta).
\label{coskp1sol}
\end{align}
The quantities used in the above equation are defined below.
$x_1,x_2,x_3$ are the roots of the following cubic equation 
(in the order $x_1 <x_2 
<x_3$\footnote{Ref.~\cite{cho2019brd} uses a different ordering of the roots
$x_2 < x_3 <x_1$.})
\begin{widetext}
\begin{align}
& -2 \delta_1 L S_1\left(\delta_1-\delta_2\right) x^3 -\left[L^2\left(\delta_1-\delta_2\right)^2+2 \delta_2 L \Sigma_2 S_2\left(\delta_2-\delta_1\right)+\delta_1^2 S_1^2+2 \delta_1 \delta_2 \Sigma_1 S_1 S_2+\delta_2^2 S_2^2\right] x^2 \nn\\
&+2 \delta_2 S_2\left[L \Sigma_1\left(\delta_2-\delta_1\right)+\Sigma_2\left(\delta_1 S_1+\delta_2 \Sigma_1 S_2\right)\right] x-\delta_2^2 S_2^2\left(\Sigma_1^2+\Sigma_2^2-1\right)=0.   
\end{align}
\end{widetext}
$\Sigma_{1}, \Sigma_{2}$, $\delta_{1}$, and $\delta_{2}$  are constant quantities
 defined as
\begin{align}
\Sigma_{1} & \equiv \frac{\vec{S}_1\cdot \vec{S}_2}{S_1 S_2}-\frac{L}{S_2} \frac{\delta_{1}-\delta_{2}}{\delta_{2}} \frac{\vec{L}\cdot \vec{S}_1}{L S_1}   ,  \label{const_1} \\
\Sigma_{2} &  \equiv \frac{\vec{L}\cdot \vec{S}_2}{L S_2}+ \frac{\delta_1 S_1}{\delta_2 S_2}  \frac{\vec{L}\cdot \vec{S}_1}{L S_1}  ,     \label{const_2}  \\ 
\delta_a &  \equiv  2 ~\nu~ \sigma_a, ~~~~~~~(a= 1, 2).
\end{align}
It is easy to rewrite $\Sigma$'s in terms of
other constants of motion as (see Eqs.~A8 and A9 of Ref.~\cite{Tanay:2021bff})
\begin{align}
\Sigma_1 &=\frac{\sigma_2\left(J^2-L^2-S_1^2-S_2^2\right)-2 S_{\text {eff }} \cdot L}{2 \sigma_2 S_1 S_2} ,  \\
\Sigma_2  &=\frac{S_{\mathrm{eff}} \cdot L}{\sigma_2 L S_2},
\end{align}
which goes on to show that the $\Sigma$'s are also constants 
of motion.
We still need more description to 
fully explicate the solution given in
Eq.~\eqref{coskp1sol}. 
\begin{align}
\Upsilon(t) & = \frac{\sqrt{A\left(x_{3}-x_{1}\right)}}{2}\left(\alpha+\frac{v +e \sin v}{c^{2} {l}^{3}}\right),       \\
\beta  & = \sqrt{\frac{x_{2}-x_{1}}{x_{3}-x_{1}}}   ,  
\end{align}
\begin{align}
\alpha  & =  \frac{ \pm 2}{\sqrt{A\left(x_{3}-x_{1}\right)}}         \nn \\  
        &   F\left(\arcsin \sqrt{\frac{\cos \kappa_{1}(t=0)-x_{1}}{x_{2}-x_{1}}}, \beta\right)  ,
\label{alphasol}\\
 A  & = 2  l s_1 \delta_1(\delta_2-\delta_1) ,           \\
v & = 2 \arctan \left(\sqrt{\frac{1+e}{1-e}} \tan \frac{u}{2}\right) ,   ~~~\text{or}      \label{anomaly_1}     \\    
    &  =  u + 2 \arctan \left(     \frac{\beta_{e} \sin u}{ 1 - \beta_e  \cos u}   \right) ,       \label{anomaly_2}  \\
    \beta_e   &  = \frac{e}{1 + (1  -   e^2)^{1/2}}   ,    \\
n &\left(t-t_{0}\right)=u-e \sin u   ,        \label{Kepler_eqn}   \\
  n  & =    \left( - 2 h \right)^{3/2}    ,      \label{mean_motion_newt} \\
  e   & =  \sqrt{ 1 + 2 h l^2},                \label{ecc_newt}      \\ 
  F(\phi_p, k)  &  \equiv \int_0^{\phi_p} \frac{\mathrm{d} \theta}{\sqrt{1-k^2 \sin ^2 \theta}}   ,      \label{jacobi-1}   \\
  \operatorname{sn}(F, k)  &  \equiv \sin (\operatorname{am}(F, k)) \equiv \sin \phi_p     \label{jacobi-2} .
\end{align}
Note that Eqs.~\eqref{anomaly_1}, \eqref{Kepler_eqn}, \eqref{mean_motion_newt}, and 
\eqref{ecc_newt} give the true anomaly,
Kepler equation, mean motion and the eccentricity,
 respectively, all at the Newtonian level. 
 Also, it is important to mention
  that Eq.~\eqref{anomaly_2} is preferred over Eq.~\eqref{anomaly_1}
because its RHS does not have the discontinuity due to 
$\arctan$, the way the RHS of Eq.~\eqref{anomaly_1} has. This renders the latter
not so useful beyond the timespan of one orbital period.
In Eqs.~\eqref{jacobi-1} and \eqref{jacobi-2}, $F$ is the incomplete
elliptic integral of the first kind, whereas sn and am are the Jacobi sin and amplitude functions respectively.
Finally, the $+$ sign in Eq.~\eqref{alphasol} is chosen if  ${d \cos \kappa_1}/{dt}>0$ at initial time,
 otherwise we choose the $-$ sign.

The cosine of the other two remaining angles $\kappa_2$ and
 $\gamma$ between the various angular momenta 
can be easily had from the solution for $\cos \kappa_1$ 
(given in Eq.~(\ref{coskp1sol})),
supplemented by Eqs.~\eqref{const_1} and \eqref{const_2}.
This finally leads to
\begin{align}
\cos \gamma &={\Sigma}_{1}+\frac{L}{S_{2}} \frac{\delta_{1}-\delta_{2}}{\delta_{2}} \cos \kappa_{1},   \label{otherangles_0} \\
\cos \kappa_{2} &={\Sigma}_{2}- \frac{ \delta_1 S_{1}}{ \delta_2 S_{2}} \cos \kappa_{1}     .
\label{otherangles}
\end{align}

\subsection{Solution for $\vec{L}$}

For the solution of $\vec{L}$, we will use an inertial frame (call it IF),
  whose $z$-axis is aligned with $\vec{J}$; see Fig.~\ref{fig:1}.
  In this frame, $\vec{L}$ has a polar angle
$\theta_L$ and an azimuthal angle $\phi_L$. 
Because $dL/dt = 0$, we need to determine only these two angles to determine $\vec{L}$.
Since
\begin{align}
\cos \theta_L  &=  \frac{\vec{L} \cdot \vec{J}}{ L J} 
              =   \frac{L^2 + L S_1 \cos \kappa_1 + L S_2 \cos \kappa_2 }{L J}  ,
\end{align}
this means that we can construct $\cos \theta_L$ as a function of time from our
already-constructed solutions for the cosines of $\kappa_1, \kappa_2$ and $\gamma$
in Sec.~\ref{mutual_ang_sec}.

\figA

The  solution for the azimuthal angle $\phi_L$ of $\vec{L}$ in  the
IF and its method of derivation closely follows Sec.~III-B of Ref.~\cite{cho2019brd}.
So without presenting the derivation, we again directly present the solution 
which reads
\begin{align} 
\phi_L(t)  -  \phi_{L_0}  =  \mathcal{F}(t),               \label{phi_L_sol}
\end{align}
where the function $  \mathcal{F} (t)$ is
\begin{widetext}
\begin{align}
 \mathcal{F} (t)=\frac{2}{\sqrt{A\left(x_{3}-x_{1}\right)}}\left(\frac{\beta_{1} \Pi\left(\frac{x_{1}-x_{2}}{\alpha_{1}+x_{1}}, \operatorname{am}(\Upsilon, \beta), \beta\right)}{\alpha_{1}+x_{1}}-\frac{\beta_{2} \Pi\left(\frac{x_{1}-x_{2}}{\alpha_{2}+x_{1}}, \operatorname{am}(\Upsilon, \beta), \beta\right)}{\alpha_{2}+x_{1}}\right),
\end{align}
with
\begin{align}
&\alpha_{1}=-\frac{\delta_{2}\left(j+l+s_{2} {\sigma}_{2}\right)}{s_{1}\left(\delta_{1}-\delta_{2}\right)}  , \\
&\alpha_{2}=-\frac{\delta_{2}\left(-j+l+s_{2} {\sigma}_{2}\right)}{s_{1}\left(\delta_{1}-\delta_{2}\right)} ,\\
&\beta_{1}=-\delta_2\frac{{l}^{2} \delta_{1}+{j}^{2} \delta_{2}+{s}_{1}\left(\delta_{1}-\delta_{2}\right)\left({s}_{1}+{s}_{2} {\sigma}_{1}\right)+{l} {s}_{2} \delta_{1} {\sigma}_{2}+{j}\left[{l}\left(\delta_{1}+\delta_{2}\right)+{s}_{2} {\sigma}_{2} \delta_{2}\right]}{2 {s}_{1}\left(\delta_{1}-\delta_{2}\right)}, \\
&\beta_{2}=-\delta_2\frac{{l}^{2} \delta_{1}+{j}^{2} \delta_{2}+{s}_{1}\left(\delta_{1}-\delta_{2}\right)\left({s}_{1}+{s}_{2} {\sigma}_{1}\right)+{l}{ s}_{2} \delta_{1} {\sigma}_{2}-{j}\left[{l}\left(\delta_{1}+\delta_{2}\right)+{s}_{2} {\sigma}_{2} \delta_{2}\right]}{2 {s}_{1}\left(\delta_{1}-\delta_{2}\right)}.
\end{align}
$\phi_{L_0}$ is an integration constant to be determined
 by substituting $t = 0$ in Eq.~\eqref{phi_L_sol}\footnote{The corresponding 
equations in Ref.~\cite{cho2019brd} (Eqs.~(3.28)) have typos.}.

\subsection{Solution for $\vec{S_1}$ and $\vec{S_2}$}

In Ref.~\cite{cho2019brd}, the full solution of $\vec{S}_1$ was not presented 
in the same way as for $\vec{L}$. 
However, we can construct the solution for $\vec{S}_1$ in a way that
parallels very closely the method that Ref.~\cite{cho2019brd} adopted to
construct the solution of $\vec{L}$\footnote{$\vec{L}$ solution was constructed in Ref.~\cite{cho2019brd}
by introducing an ``$\vec{L}$-centered'' non-inertial frame, whose $z$-axis was aligned 
with $\vec{L}$. To construct the $\vec{S}_1$ solution using this method, we need to
introduce an ``$\vec{S}_1$-centered'' non-inertial frame whose $z$-axis is aligned with $\vec{S}_1$.}.
So, again we merely present the solution without derivation.

Just like $\vec{L}$, $\vec{S}_1$ is described by its polar and azimuthal angles ($\theta_{S1}$ and
$\phi_{S1}$ respectively) because its magnitude
is fixed. For the polar angle, we have just like before
\begin{align}
\begin{aligned}
\cos \theta_{S1} &=\frac{\vec{S}_1 \cdot \vec{J}}{S_1 J}=\frac{S_1^2+\vec{L} \cdot \vec{S}_1+\vec{S}_1 \cdot \vec{S}_2}{S_1 J} \\
&=\frac{S_1^2+L S_1 \cos \kappa_1+ S_1 S_2 \cos \gamma }{S_1  J},
\end{aligned}
\end{align}
which means that $\cos \theta_{S1}$ can be had from our previously constructed solutions of 
cosines of $\kappa_1, \kappa_2$ and $\gamma$.

The solution for the azimuthal angle $\phi_{S1}(t)$ of $\vec{S}_1$ in the IF is given by
\begin{align}
\phi_{S_1}(t)   -  \phi_{S_{10}}= \mathcal{G}(t)     ,    \label{S1_angle}
\end{align}      
where the function $\mathcal{G}(t)$ is
\begin{align}
\mathcal{G} (t) =  \frac{2}{\sqrt{A\left(x_{3}-x_{1}\right)}}\left(\frac{\beta_{1s1} \Pi\left(\frac{x_{1}-x_{2}}{\alpha_{1s1}+x_{1}}, \operatorname{am}(\Upsilon, \beta), \beta\right)}{\alpha_{1s1}+x_{1}}-\frac{\beta_{2s1} \Pi\left(\frac{x_{1}-x_{2}}{\alpha_{2s1}+x_{1}}, \operatorname{am}(\Upsilon, \beta), \beta\right)}{\alpha_{2s1}+x_{1}}\right)  +  \frac{v+e \sin v}{c^2 l^3}  j  \delta_2 ,
\end{align}
with
\begin{align}
&\alpha_{1s1}=\frac{\delta_{2}\left(-j+s_1+s_{2} {\sigma}_{1}\right)}{l \delta_1}   ,\\
&\alpha_{2s1}=\frac{\delta_{2}\left(j+s_1+s_{2} {\sigma}_{1}\right)}{l \delta_1}   ,\\
&\beta_{1s1}=-\delta_2\frac{{l}^{2} \delta_{1}+\left({s}_1 (\delta_1 -\delta_2) +{j} \delta_2\right)\left(-{j} +{s}_1 +{s}_2 {\sigma}_1\right)+ {l} {s}_2 \delta_1 {\sigma}_2}{2 {l} \delta_1}, \\
&\beta_{2s1}=\delta_2\frac{-{l}^{2} \delta_{1}+\left(-{s}_1 \delta_1 + ({j}+{s}_1) \delta_2\right)\left({j} +{s}_1 +{s}_2 {\sigma}_1\right)- {l}{ s}_2 \delta_1 {\sigma}_2}{2 {l} \delta_1} .
\end{align}
\end{widetext}
$ \phi_{S_10}$ is an integration constant to be determined by substituting 
$t=0$ in Eq.~\eqref{S1_angle}. This completes the solution for $\vec{S}_1$.
Now $\vec{S}_2$ can also be easily had from
\begin{align}           \label{S_2_sol}
\vec{S}_2  =  \vec{J} - \vec{L} -\vec{S}_1   .   
\end{align}

\subsection{Solution for $\vec{R}$}    \label{r-vector-sol}

Specification of $\vec{R}$ can be broken down into its magnitude $R$ and its
azimuthal angle (phase) $\phi$ in a precessing
 plane that is perpendicular to $\vec{L}$.
The solution for $\vec{R}$ proceeds along the same lines as in Ref.~\cite{cho2019brd},
but is  somewhat modified due to the inclusion of the 1PN term in the Hamiltonian.
So, unlike previous sub-sections, we will present 
the derivation of the solution for $\phi$. 
Since  the magnitude of $\vec{R}$ has already been
worked out in the literature, we will simply 
state the relevant results
without presenting the derivation.

Let us first focus on the magnitude $r$ of the scaled position vector
$\vec{r} =  \vec{R}/(G M)$.
The 1PN and 2PN contributions (without spins)
can be found in Ref.~\cite{1985AIHS...43..107D}, whereas the 1.5PN contribution (with 
the 1PN effects ignored) can  be found in Ref.~\cite{Gopakumar:2011zz}. Combining these results together,
we have up to 1.5PN order the following quasi-Keplerian parametric solution for $r$
\begin{align}         \label{r_newt}
\begin{aligned}
&r=a_r\left(1-e_r \cos u\right)   , \\
& l' \equiv n\left(t-t_0\right)=u-e_t \sin u  , 
\end{aligned}
\end{align}
with $u$ and $l'$ standing for the eccentric and the mean anomalies, respectively.
The other constants that comprise the above solution are 
{
\allowdisplaybreaks[1]
\begin{align}
a_r &=-\frac{1}{2 h}\left(1  -\frac{1}{2} (\nu - 7)\frac{h}{c^2}   -2   \frac{ \Ef}{l^2}  \frac{h}{c^2}\right)  ,    \label{ar_PN}  \\
e_r^2 &=   1  +  2 h l^2  - 2(6-\nu) \frac{h}{c^2} - 5(3-\nu)  \frac{h^2 l^2}{c^2}    \nn \\
      &  +  8  \left(1+ h l^2\right) \frac{   \Ef }{l^2} \frac{h}{c^2}, \\
n &=(-2 h)^{3 / 2}   \left(  1 +  \frac{2 h}{8 c^2} (15 - \nu) \right)   ,   \label{mean_motion_1.5PN}    \\
e_t^2 &=  1 + 2 h l^2  + 4 (1-\nu) \frac{h}{c^2} + (17- 7 \nu)  \frac{h^2 l^2}{c^2}    \nn \\   
     &   + 4 \frac{ \Ef}{l^2} \frac{h}{c^2},
\end{align}
}
where the following definitions have been used
\begin{align}
\vec{s}_{\text{eff}}    & \equiv  \delta_1 \vec{s}_1   +   \delta_2 \vec{s}_2,   \\
\Ef   &  \equiv  \vec{s}_{\text{eff}}   \cdot \vec{l} .  
\end{align}
Recall that $\Ef$  is a constant since its PB with $H$ vanishes.
This definition, along with Eq.~\eqref{define_Seff}
 implies the following relation
\begin{align}
\vec{s}_{\text{eff}}   \equiv  \frac{2 }{ G M^2} \vec{S}_{\text{eff}} .
\end{align}

With the magnitude of $\vec{r}$ being  taken care of,
  we can now move on to determine its 
phase in a plane perpendicular to $\vec{l}$.
The scaled-time derivative (wrt. $t$) of the scaled position vector $\vec{r}$ is
\begin{align}
&  \frac{d \vec{r}}{dt}   = \pb{\vec{r}, H}   G M   \\
 =& \vec{p}  +   \frac{1}{ 2 c^2  r^3}  \left(   - 2 \nu ( \vec{p} \cdot \vec{r}) \vec{r} - (6+2 \nu)  r^2  \vec{p}      \right.  \nn \\
  &   \left.  + (3 \nu -1)  p^2  r^3 \vec{p}  - 2   \vec{r} \times \vec{s}_{\text{eff}}     \right),
\end{align}
which when solved perturbatively gives the following 
expression for $\vec{p}$
\begin{align}           \label{pvec_perturb}
  \begin{split}
    \vec{p} = \frac{d \vec{r}}{dt}
    + \frac{1}{c^2} \Bigg[&    \frac{\vec{r} \times \vec{s}_{\text{eff}}   +  \nu (\vec{p} \cdot \vec{r} ) \vec{r}  }{r^3}  \\
    &{}+ \frac{d \vec{r}}{dt} \left( \frac{p^2}{2} - \frac{3 p^2
        \nu}{2} + \frac{3+\nu}{r} \right) \Bigg]
  \end{split}
\end{align}

At this point, we introduce a non-inertial frame (call it NIF) whose
$x, y$ and $z$ axes are along $\vec{J} \times \vec{L}$,
$\vec{L} \times (\vec{J} \times \vec{L})$ and $\vec{L}$, respectively.
The Euler matrix which when multiplied with a column
consisting of a vector's components in the IF, gives its 
components in the NIF is
\begin{align}          \label{Euler}
E_L =  \left(\begin{array}{ccc}
-\sin \phi_L & \cos \phi_L & 0 \\
-\cos \theta_L \cos \phi_L & -\cos \theta_L \sin \phi_L & \sin \theta_L \\
\cos \phi_L \sin \theta_L & \sin \theta_L \sin \phi_L & \cos \theta_L
\end{array}\right) .
\end{align}
In the NIF, the various vector components are
\begin{align}       \label{l_NIF}
\vec{l}=\left[\begin{array}{l}
0 \\
0 \\
l
\end{array}\right]_n
\end{align}
\begin{align}              \label{s1_NIF}
\vec{s}_1=s_1\left[\begin{array}{c}
\sin \kappa_1 \cos \xi \\
\sin \kappa_1 \sin \xi   \\
\cos \kappa_1
\end{array}\right]_n
\end{align}
\begin{align}             \label{s2_NIF}
\vec{s}_2=s_2\left[\begin{array}{c}
\sin \kappa_2 \cos \left(\xi+ \psi\right) \\
\sin \kappa_2 \sin \left(\xi+ \psi\right) \\
\cos \kappa_2
\end{array}\right]_n
\end{align}
\begin{align}            \label{r_NIF}
\vec{r}=\left[\begin{array}{c}
r \cos \phi \\
r \sin \phi \\
0
\end{array}\right]_n
\end{align}
Note that with the aid of Eqs.~\eqref{s1_NIF} and \eqref{s2_NIF},
we have actually defined the angles $\xi$ and $ \psi$, that is to say, 
$\xi$ and $\xi + \psi$ are azimuthal angles of $\vec{s}_1$ and $\vec{s}_2$
in the NIF. 
Implicit in Eq.~\eqref{r_NIF} is the definition that $\phi$ is the
azimuthal angle of $\vec{r}$ in the NIF.
Also, the subscript ``$n$'' after the above columns
denotes that the components have been written in the NIF.

Being cognizant of the fact that the time
 derivative of a vector changes (as a geometrical object)
with a change of the frame in which it is viewed (see Sec.~4.9 of Ref.~\cite{goldstein2013classical}),
 we mention here that all time derivatives
of vectors in this paper are to be viewed in the IF. 
With that, we now wish to express the components of
$d \vec{r}/dt$ in the NIF. We do so 
by first writing the $\vec{r}$-components in the IF (using
Eq. \eqref{r_NIF} and the inverse of $E_L$ of Eq.~\eqref{Euler}), 
taking derivatives of the components with respect to $t$,
 and then transforming the components of 
 this time derivative to the NIF (using $E_L$).
The end result is
\begin{align}        \label{rdotvec}
\dot{\vec{r}}=\left[\begin{array}{c}
\dot{r} \cos \phi-r \sin \phi\left(\dot{\phi}_L \cos \theta_L+\dot{\phi}\right) \\
\dot{r} \sin \phi+r \cos \phi\left(\dot{\phi}_L \cos \theta_L+\dot{\phi}\right) \\
r\left(-\dot{\phi}_L \sin \theta_L \cos \phi+\dot{\theta}_L \sin \phi\right)
\end{array}\right]_n
\end{align}

Now we want to express $\vec{l} = \vec{r} \times \vec{p}$ 
in component form in the NIF. The components of $\vec{l}$ and $\vec{r}$ 
in the NIF are already given by Eqs.~\eqref{l_NIF} and \eqref{r_NIF}.
For $\vec{p}$, we start with Eq.~\eqref{pvec_perturb}, where
we use Eqs.~\eqref{s1_NIF} and \eqref{s2_NIF} for expressing
 $\vec{s}_{\mathrm{eff}}$, and Eq.~\eqref{rdotvec} for $d \vec{r}/dt$.
With all this, the third 
component of the relation $\vec{l}   =  \vec{r} \times \vec{p}$
 when written out gives us  (with $\epsilon \equiv  1/c^2 $ and $h \equiv H/\mu$)
\begin{widetext} 
\begin{align}
  \frac{d \phi}{dt}  =  -\frac{2( \epsilon  s_1  \delta_1  \cos \kappa_1 + \epsilon  s_2 \delta_2  {\cos} \kappa_2 + l r )}{ r^2  \left(-2 \epsilon (3+ \nu ) + \left(-2+\epsilon (-1+3 \nu)\left( ( \vec{p} \cdot \hat{r} )^2 + \frac{l^2}{r^2}\right)\right) r  \right)}-  {\cos}\theta_L \frac{d \phi_L }{dt},     \label{phi_dot_temp}
\end{align}
We mention that to arrive at Eq.~\eqref{phi_dot_temp} we eliminated 
$p^2$ in favor of $\vec{p} \cdot \hat{r}$ (or $\vec{p} \cdot \vec{r}$) 
 with the aid of 
\begin{align}
p^2  = (\vec{p} \cdot \hat{r})^2 +  \frac{l^2}{r^2}    ,   \label{p_sqrd_eqn}
\end{align}
where $\vec{p} \cdot \hat{r}$ is given by 
(see Eq.~(36) of Ref.~\cite{Tanay:2020gfb})
\begin{align}       \label{pdn}
 \vec{p} \cdot \hat{r}  =    \pm    \left(\left(2 h  -  \epsilon (3 \nu - 1)  h^2\right)+\frac{2(1+\epsilon  (4 - \nu) h)}{r}+\frac{\left(-l^2+\epsilon  (6+\nu) \right)}{r^2}+\frac{-\epsilon \nu \left(  l^2 + 2  \Ef / \nu \right)}{r^3}\right)^{1 / 2} .        
\end{align}
A detailed discussion on how to choose the correct sign has been
relegated to Appendix~\ref{sign_of_pdn}.

We want to rewrite Eq.~\eqref{phi_dot_temp} so that the only time-varying 
quantities on the RHS are $\cos \kappa_1$ and $r$. 
Therefore, we need to eliminate $\vec{p} \cdot \hat{r}, d \phi_L/dt$, $\cos \theta_L$, and 
$\cos \kappa_2$.
To do so, we make use of
Eqs.~\eqref{otherangles}, \eqref{pdn} and the
relations
\begin{align}
   \vec{L} \cdot \vec{J}   &  =  J L \cos \theta_L   = \vec{L} \cdot (\vec{L} + \vec{S}_1 + \vec{S}_2) =L^2 + L S_1 \cos \kappa_1  +  L  S_2 \cos \kappa_2  ,     \label{LJ}   \\
\frac{d \phi_L }{dt}   &   =  \frac{1}{c^2 r^3}\left(\frac{\beta_1}{\cos \kappa_1+\alpha_1}-\frac{\beta_2}{\cos \kappa_1+\alpha_2}\right)   .    \label{phi_L_dot}
\end{align}
Eq.~\eqref{phi_L_dot} has already been derived in Ref.~\cite{cho2019brd}, see its Eq.~(3.27). Finally, Eq.~\eqref{phi_dot_temp} becomes
\begin{align}
\frac{d \phi}{dt}  &=    \frac{ \epsilon }{r^3}  \left(\frac{\beta_{1}}{\cos \kappa_1+\alpha_{1 }}    +   \frac{\beta_{2 }}{\cos \kappa_1+\alpha_{2 }}\right)    \nn  \\
 &  -\frac{l \epsilon^2(-1+3 \nu)\left(2 \Ef +{l}^2 \nu\right)}{2 r^5}+\frac{l \epsilon^2\left(-6+17 \nu+3 \nu^2\right)}{2 r^4}+\frac{l\left(2-h^2 \epsilon^2(1-3 \nu)^2+2 h \epsilon(-1+3 \nu)\right)}{2 r^2}   \nonumber  \\ 
&  -\frac{\epsilon\left(- \Ef + l^2\left(4 + 4 h \epsilon-2 \nu-13 h \epsilon \nu+3 h \epsilon \nu^2+\delta_1+\delta_2\right)+ l s_2 \delta_2 \Sigma_2\right)}{l r^3},    \label{phi_dot}  \\
&  \equiv   \frac{ \epsilon }{r^3}  \left(\frac{\beta_{1}}{\cos \kappa_1+\alpha_{1 }}    +   \frac{\beta_{2 }}{\cos \kappa_1+\alpha_{2 }}\right)    +   \frac{A_5}{r^5}    +   \frac{A_4}{r^4}    +     \frac{A_3}{r^3}  +     \frac{A_2}{r^2}  .     \label{As_defined}    
\end{align}
which when integrated gives
\begin{align}        \label{phi_sol}
\phi - \phi_0   =   \frac{2}{\sqrt{A\left(x_3-x_1\right)}}\left(\frac{\beta_{1} \Pi\left(\frac{x_1-x_2}{\alpha_{1}+x_1}, \operatorname{am}(\Upsilon, \beta), \beta\right)}{\alpha_{1}+x_1}  + \frac{\beta_{2} \Pi\left(\frac{x_1-x_2}{\alpha_{2}+x_1}, \operatorname{am}(\Upsilon, \beta), \beta\right)}{\alpha_{2} + x_1  }\right)   +   {A_5}{\mathcal{R}_5}    +   {A_4}{\mathcal{R}_4}    +     {A_3}{\mathcal{R}_3}  +     {A_2}{\mathcal{R}_2}   
\end{align}
\end{widetext}
where $\phi_0 $ is an integration constant to be determined by substituting
$t = 0$ in the above equation.
Note that Eq.~\eqref{As_defined} defines $A_2, A_3, A_4,$ and $A_5$.
 Also, $\mathcal{R}_j$ is the indefinite integral of 
$r^{-j}$ with respect to $t$. 
For brevity, we don't write out $\mathcal{R}_j$'s explicitly
 but they can be easily had
from the 1.5PN accurate expressions
\begin{align}         
&     r  =   a_r   \left(1-e_r \cos u\right)   , \\
&       n\left(t-t_0\right)=u-e_t \sin u  ,
\end{align}
where $a_r, e_r, n$, and $e_t$ are defined
in Sec.~\ref{r-vector-sol}.
Note that we have kept more than necessary PN accuracy in Eq.~\eqref{phi_dot}
and by extension, Eq.~\eqref{phi_sol} just for
the purposes of illustration;
any additive term containing $\epsilon^q$ with $q \geq 2$ can be dropped.

\subsection{Solution for $\vec{P}$}

We present the determination of $\vec{P} = \mu \vec{p}$ in a step-by-step
algorithmic fashion
\begin{enumerate}
\item Solution of $r$ as a function of time has been 
given in Sec.~\ref{r-vector-sol}. This, when used in
Eq.~\eqref{pdn}, combined with the discussion of Appendix~\ref{sign_of_pdn},
gives us $\vec{p} \cdot \hat{r}$, which when 
further combined with Eq.~\eqref{p_sqrd_eqn} 
lets us determine $p^2$ or $p$. With this, the magnitude of $\vec{p}$
is in our hands.
What now remains is to determine the azimuthal angle of $\vec{p}$ in the NIF.
\item Next we compute the angle $\phi_{\text{offset}}$, which we
define as
\begin{align}
\begin{aligned}         \label{p_azimuth}
\phi_{\text{offset}}   & =   \arcsin \frac{L(t)}{r(t) p(t)}           &~~~~~\text{if~}\vec{p} \cdot \hat{r} > 0 ,    \\
\phi_{\text{offset}}    &  =  \pi -   \arcsin \frac{L(t)}{r(t) p(t)}   &~~~~~\text{if~}\vec{p} \cdot \hat{r} < 0 .
\end{aligned}
\end{align}
\item Now it is a simple matter to see that
the azimuthal angle that $\vec{p}$ makes with the $x$-axis of the NIF is 
$\phi + \phi_{\text{offset}} $, where $\phi $ is the azimuthal angle of  $\vec{r}$ in the
 NIF and $\phi_{\text{offset}} $ is the relative azimuthal angle between $\vec{r}$ and $\vec{p}$. This completes the specification of $\vec{P}$.
\end{enumerate}

\section{Action-angle based solution}
\label{aasol}

We devote this section to explaining our construction of an alternate 
AA-based solution to the 1.5PN BBH system. We will present in a step-by-step
algorithmic fashion.

 1. The phase space of the 1.5PN system is 10 dimensional
(since spin magnitudes are constants) and has five mutually
 commuting constants, resulting in an integrable system. These constants are $\vec{\mathcal{C}} =
\left\{  J , J_{z}, L , H, \Eff \right\}$. We thus have five action variables 
$\vec{\mathcal{J}}=\left\{   \mathcal{J}_{1} = J , \mathcal{J}_{2}  = J_{z},
\mathcal{J}_{3} =  L, \mathcal{J}_{4}, \mathcal{J}_{5}
\right\}$, which can be found in Refs.~\cite{Tanay:2020gfb} and \cite{Tanay:2021bff};
their notations and definitions are somewhat different from those in this 
paper\footnote{$\mathcal{J}_{4}$ was obtained by using 
Damour and Sch{\"a}fer's PN extension of the 
Sommerfeld radial action complex contour
integral \cite{damour1988}, whereas $\mathcal{J}_{5}$ was obtained 
using the method of fictitious phase-space variables~\cite{Tanay:2021bff}.}.

2. Under the real-time evolution of the BBH system, the angles change as
$ \Delta \theta_i  =  \omega_i t'$, $t' = t G M$ being the physical time;
$\omega_i$'s are naturally called the frequencies of the system and are constants
since they are functions of the action variables, which are themselves constants.
Sec.~VI-A of Ref.~\cite{Tanay:2021bff} shows  how to compute the five frequencies.
The process mainly consists of constructing the Jacobian matrix of the actions 
with respect to the commuting constants, inverting this Jacobian matrix, and the 
row corresponding to the Hamiltonian (one of the commuting constants) contains 
 the five frequencies.

3. Let all the phase space variables be collectively denoted by $\vec{V} =\left\{
\vec{R},\vec{P}, \vec{S}_1, \vec{S}_2
\right\}$ with their initial values being $\vec{V}_0 \equiv \vec{V}(t'=0)$.
Suppose that $\vec{V}_0$ represents the state of the system 
at $t' = 0$, and we want to obtain $\vec{V}(t')$ at any non-zero time $t'$.

4. At $t' = 0$, assign all the angle variables of the system to have the value 
equal to 0; $\vec{\theta}(t' = 0) = \vec{0} $. Then at the final time $t'$,
the angles of the system would have become $\vec{\theta} = \vec{\omega}\times t'$.
The problem of finding the state of the system at $t'$ now becomes that of
increasing all the angle variables by $\vec{\omega}\times t'$.

5. As shown in Sec VI-B of Ref.~\cite{Tanay:2021bff}, the angle $\theta_i$ can be 
increased by a certain amount if we flow under the corresponding action ${\mathcal{J}}_i$
by the same amount. Therefore the problem becomes that of flowing under the actions
$\mathcal{J}_i$ by amounts $\omega_i t'$ (starting from the $\vec{V}_0$ configuration).
The order of flows does not matter because just like all the members of $\vec{C}$,
all the actions mutually commute among themselves too.

6. Because the flow equation under the actions reads
\begin{align}
\begin{aligned}
\frac{d \vec{V}}{d \lambda} &=\left\{\vec{V}, \mathcal{J}_i(\vec{C})\right\} 
=\left\{\vec{V}, C_j\right\} \frac{\partial \mathcal{J}_i}{\partial C_j} ,
\end{aligned}
\end{align}
a flow under an action $\mathcal{J}_i$ by $\Delta \lambda_i$
can be achieved by flowing under all the commuting constants $C_j$'s
by respective amounts $ ( \partial \mathcal{J}_i/\partial C_j  ) \Delta \lambda_i$.
Again, we can flow under the commuting constants $C_j$'s in any order, since
they all mutually commute.
This finally breaks down our problem to that of flowing under all five $C_j$'s
by certain amounts, whose closed-form solution is discussed
in Appendix \ref{commuting_const_flow}.

Now it should be clear as to why one of the inputs to the
 AA-based solution is the Standard 
Solution that was presented in Sec.~\ref{hmsol}. 
This is so because one of the commuting constants
is the Hamiltonian itself and in 
Sec.~\ref{hmsol}, we presented the solution of
the flow under this commuting constant. This also happens 
to be the solution of the system since the system's evolution
equations are same as the equations governing the
flow under the Hamiltonian.
Nevertheless, the AA-based solutions appears 
to have the advantage to be pushed
to 2PN using canonical perturbation theory
which is a perturbation technique designed around the action-angle framework \cite{goldstein2013classical}.

\section{Software package and comparison of the solutions}
\label{cmp}

With this article, we are releasing a public \textsc{Mathematica} package \texttt{BBHpnToolkit},
which accomplishes the following objectives~\cite{MMA1}.
\begin{enumerate}
\item Implements the Standard Solution of Sec.~\ref{hmsol} obtained by integrating the 
flow under the Hamiltonian.
\item Implements the action-angle-based solution of Sec.~\ref{aasol}. 
\item Provides the numerical solution of the system by numerically
integrating the EOMs.
\item Provides  numerical values of all five frequencies of the system,
 wherein by frequency we refer to the rates of change of 
the five angles (as in action-angle variables) of the system.
\end{enumerate}
For the reference of the future users of 
our \textsc{Mathematica} package, we mention herein that 
we have retained some unnecessary high PN terms in
 the coded expressions which implement the two 
analytical solutions in the package.
The full account is given in Appendix~\ref{MMA_accuracy}.
We devote this section to
investigating the PN accuracy of the two analytical solutions with respect to the one obtained by
direct numerical integration of the EOMs. 
In this section, we will refer to both these
 two analytical solutions collectively
as just ``analytical solutions'', for these two 
analytical solutions agree with each much better than they do with the
numerical one.

\begin{figure*}[t!]         
    \centering
    \begin{subfigure}[t]{0.45\textwidth}      
        \centering
        \includegraphics[height=2.2in]{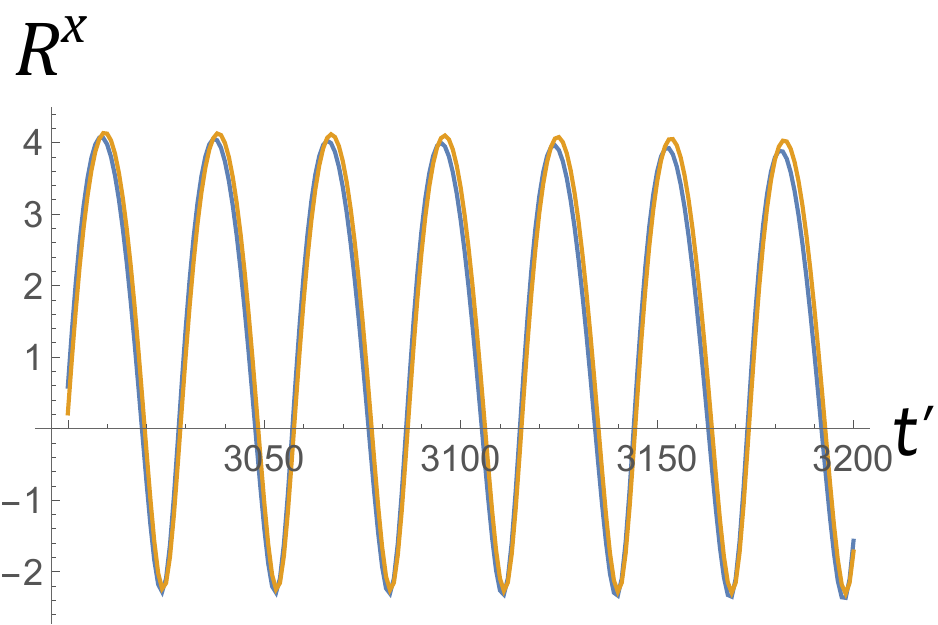}
        \caption{}
    \end{subfigure}%
    ~ 
    \begin{subfigure}[t]{0.45\textwidth}
        \centering
        \includegraphics[height=2.3in]{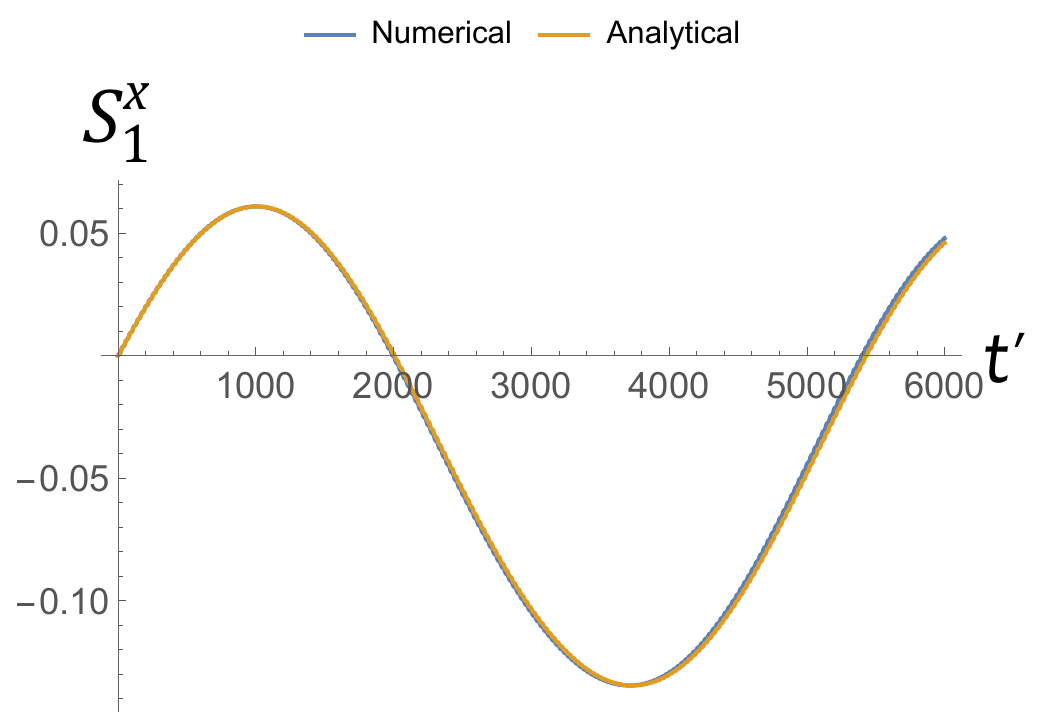}
        \caption{}
    \end{subfigure}
    \caption{   Comparison of the analytical solutions with the numerical one. For a system with $(m_1,m_2)=(  5/2, 1)$ and the initial values of the phase-space variables being 
    $\vec{R}=(2,$ $2,2),$ $~\vec{P}=(1/2,$ $-1/2, 1/3),~$  $\vec{S_1}$ $=\sqrt{\epsilon}$ $~(0,1,$ $1),~\vec{S}_2$ $=\sqrt{\epsilon}~(1, $ $-3/10, $ $0)$.
     Subfigures (a) and (b) show evolution of $x$-component of $\vec{R}$  and  $\vec{S}_1$, respectively.
    We choose $\epsilon=0.003$ for (a) and $\epsilon=0.01$ for (b). All this results in a  Newtonian-orbital time period of $T_N \sim 29$ for both (a) and (b), and
    the  PN parameter $\sim 0.0036$  for (a) $\sim 0.012$ for (b) respectively. Throughout we keep $G = 1$. }     \label{plots}   
\end{figure*}

\figB

We first begin by displaying the plots of the analytical solutions along with the
numerical one in Fig.~\ref{plots}. Plots corresponding to the two analytical solutions (Standard and
AA-based solutions) cannot be distinguished visually. But they are collectively distinguishable 
from the numerical solution at late times. As displayed in the caption of this figure, 
the parameters taken for these plots do not correspond to astrophysically relevant systems.
But this is not a cause of concern because our focus in this section is to explore the 
mathematical accuracy of our solutions which are valid for  a wide set of parameters, and 
not just those which correspond to real astrophysical scenarios. Later in this section,
we will see that it is actually helpful to study these solutions under non-physical
scenarios (such as changing the value of $\epsilon \equiv 1/c^2$), to determine the PN accuracy of our 
solutions (via the method of linear fit between the error and the PN parameter).

Let us now switch to a more technical task of 
determining the PN accuracy of our 
analytical solutions against the numerical one.
We explain our comparison method with the aid of
Fig.~\ref{fig:2}. Starting from $t' = t_0 = 0$, we evolve the system numerically up to 
a certain time $t' = t_1$, such that $\vec{R}_n$ ($\vec{R}$ obtained via numerical integration)
differs by a small amount
from $\vec{R}_a$ ($\vec{R}$ obtained from the analytical solution).
By ``small amount'', we mean  an opening angle ($\zeta_3$ in Fig.~\ref{fig:2})
 of $\sim 0.5^{\circ}$ to $1^{\circ}$
between $\vec{R}_a$ and $\vec{R}_n$. The dashed circle in the figure denotes the plane
perpendicular to $\vec{L}_n$ ($\vec{L}$ obtained via 
numerical integration at $t'=t_1$).
$\vec{R}_{a \parallel}$ is the projection of $\vec{R}_a$ on this plane.
The difference between $\vec{R}_n$ and $\vec{R}_a$ is greatly exaggerated in
this figure for the purposes of illustration.
Now, even if we ignore the numerical roundoff errors, we expect 
$ \vec{R}_a  \neq   \vec{R}_n$ because of the PN truncation errors that crept in  while
deriving the analytical solutions. For this section, we force the roundoff errors to
be much smaller than the truncation errors by keeping sufficient numerical precision
while constructing the numerical solution.

Now, the $\vec{R}_a - \vec{R}_n$ can be broken down into two components: an in-plane
component $  \vec{R}_{a \parallel} - \vec{R}_n $,
 and an out-of-plane component $\vec{R}_a - \vec{R}_{a \parallel}$.
 The in-plane difference is brought about by an in-plane (perpendicular to $\vec{L}$) 
 dephasing which occurs due to the difference between the numerical and the analytical
 values of the mean motion $n$ (a kind of frequency). The out-of-plane difference is 
 brought about by the difference $\vec{L}_a - \vec{L}_n$, where $\vec{L}_a$ is the analytical $\vec{L}$.

For comparison purposes, it is convenient to
think of the in-plane and out-of-plane differences as in-plane and 
out-of-plane dephasings happening due to a difference in some
angular velocities $d \omega_{\parallel},$ and $d \omega_{\perp}$ in the two directions,
with the former being essentially the difference between the analytical and numerical
mean motions. As is suggested from Eq.~\eqref{mean_motion_1.5PN}, these frequencies are often 
expressible as a PN series. Hence, we schematically write the difference between the numerical and 
analytical frequencies as a PN series along the two directions as
\begin{align} 
d \omega_{\parallel}   & =    \xi^{p}    d \omega_{\parallel p}  +  \xi^{p+1}    d \omega_{\parallel p+1} + {\cal{O}}(\xi^{p+2}) ,   \label{errPN1}  \\
d \omega_{\perp}    &  =    \xi^{q}    d \omega_{\perp q}  +  \xi^{q+1}    d \omega_{\perp q+1} + {\cal{O}}(\xi^{q+2})   ,             \label{errPN2}
\end{align}
with $\xi$ standing for the PN parameter $G M/(c^2 R)$.
The above equations indicate that the differences in the frequencies
along the two directions occurs at PN orders $p$ and $q$. Note that unlike earlier sections,
$p$ does not stand for the magnitude of the scaled momentum, but is rather a
PN power-counting index.
It then follows that
\begin{align}
| \vec{R}_{a \parallel} - \vec{R}_n|    &  =    \xi^p  {R}_n  d \omega_{\parallel p}   t_1   ,      \\
| \vec{R}_{a \parallel} - \vec{R}_a |        &  =     \xi^q  {R}_n  d \omega_{\perp q}   t_1, 
\end{align}
where the LHS is the small arc length which subtends an angle 
$ \xi^p    d \omega_{\parallel p}   t_1$ or $ \xi^p    d \omega_{\perp p}   t_1$
a distance $R_n$ away.
Adding the vectors $\vec{R}_n  - \vec{R}_{a \parallel} $ and  $\vec{R}_{a \parallel} - \vec{R}_a$
and taking the modulus of the sum,
with the aid of the above two equations 
gives us at leading order
\begin{align}
|\vec{R}_n - \vec{R}_a|    &=     \xi^r  t_1  {R}_n    d \omega_{\parallel r}   & ~~\text{if~} p \neq q          \\ 
                           &=   \xi^r  t_1  {R}_n  \left(  d \omega_{\parallel p}^2 + d \omega_{\perp p}^2  \right)^{1/2}       & ~~\text{if~} p=q  ,    \label{peqq_case}
\end{align}
where $r = \text{min}(p,q)$\footnote{Here $\text{min}(p,q) \equiv p$ if $p<q$,
and $\text{min}(p,q)  \equiv p$ if $p=q$.}, 
which is not to be confused with 
$r \equiv |\vec{r}|$. This finally leads to
\begin{align}
\log \left(\frac{|\vec{R}_n - \vec{R}_a|}{t_1  {R}_n}   \right)  &=     r \log \xi  +     \text{constant}.       \label{testing}
\end{align}
In Eq.~\eqref{peqq_case}, use has been made
 of Pythagoras theorem since $\vec{R}_n - \vec{R}_{a \parallel} $
is perpendicular to $\vec{R}_{a \parallel} - \vec{R}_a$, 
and $\vec{R}_n - \vec{R}_a$ forms the hypotenuse.

Eq.~\eqref{testing} is the key to testing the PN accuracy of our analytical solutions against the numerical one.
It says that a linear fit between the two logarithms has a slope equal to the integer $r$. From 
Eqs.~\eqref{errPN1} and \eqref{errPN2}, we know that $r$ is also 
the PN order at which our analytical solution starts to diverge from the numerical one, 
if one focuses their attention on the frequencies associated with the dephashings.
In our numerical experiments with a handful of cases, we find $r = 2 \pm 2.5 \%$, which indicates that
the analytical solutions (both kinds) deviate from the numerical one at 2PN order, as 
far as the combined (in-plane and out-of-plane) dephasing effect goes.
To perform the linear fit associated with Eq.~\eqref{testing},
we needed the values of the PN parameter $\xi$.
We simply used $\xi = G M/(c^2  \left\langle R  \right\rangle)$,
where $\left\langle R  \right\rangle$ is the numerical 
orbit-averaged value of $R$.

Another way to compare the analytical solutions with the numerical one is to
compare the timescale $T_{N}$ of variation of the Newtonian orbit against timescale $T_D$ at which 
$\vec{R}_n - \vec{R}_a$  varies.
Assuming a relation of the form
\begin{align}
T_N  = \xi^r T_D ,
\end{align}
our objective is to determine $r$. 
The above equation gives
\begin{align}
\log T_D  = - r \log \xi + \log T_N ,      \label{testing_2}
\end{align}
A linear fit between $\log T_D$ and $\log \xi$ 
 gives us a slope of $-r$.
 For the purposes of linear-fitting,
we can approximate $T_D$ as
\begin{align}
T_D  =  \frac{2 \pi t_1}{\arccos \hat{R}_n \cdot \hat{R}_a},
\end{align}
and $\xi$ is to be approximated the same way as before. This linear fit again 
gave us $r = 2 \pm 2.5 \%$, which means that the analytical and the numerical solutions  differ at 2PN order. We finally mention that the verification
of the solutions presented here also serves as a verification for the five
action variables constructed in Refs.~\cite{Tanay:2020gfb, Tanay:2021bff}.

There is a subtle aspect of the above procedure of determining the PN accuracy of the 
analytical solutions, which needs further elaboration.
A question arises: how to vary the PN parameter $\xi \equiv G M/(c^2 R)$
to prepare the data for linear-fitting as per Eqs.~\eqref{testing}
and \eqref{testing_2}?
We put forth two options: (1) Vary the average value of $R$ (physically sensible)
(2) Vary $\epsilon = 1/c^2$ (unphysical). 
We will argue below that the latter option,
though unphysical, is the mathematically sound option and the former is not.

The multiplicative factors attached to powers 
of $\xi$ in Eqs.~\eqref{errPN1} and \eqref{errPN2}
are assumed to depend on quantities like energy, 
mass ratio etc. This comes from our
previous experience of the PN literature; for
 example see the expression of the mean motion
in Eq.~\eqref{mean_motion_1.5PN} or 
Eq.~(28-b) of Ref.~\cite{Cho:2021oai}. 
So this means that care must be taken
while selecting the sample BBH systems for
 performing the above linear fits so that all the systems which
are supposed to fall on a straight
 line have almost  the same energy and mass ratio.
 One way to achieve this is to vary $c$, 
while keeping other parameters like $m_1, m_2$ and average $R$ 
almost constant. Doing so does vary $\xi$. 
Another way to vary $\xi$ would be to 
vary the masses or the average $R$. Doing so
would also change the energy and the 
mass-ratio in general, which is something we want to avoid.
The unphysical aspect of varying the speed of light is of 
no concern to us as far as checking the PN accuracy of the
analytical solutions is concerned. This is so because our 
solutions to the BBH system are 
valid for any arbitrary positive speeds of light,
provided $\xi \ll 1$.

At this point, we would like to mention that if the above procedure 
of linear fitting to determine the PN accuracy of our solutions
(embodied in Eq.~\eqref{testing}) is 
applied to the spin vectors, we get $r \sim 1.5$, rather than 2.
However, this is no cause for alarm. The evolution 
equation for the spins due 
to only the 1.5PN part of the Hamiltonian
is ($A=1, 2$)
\begin{align}
\frac{d \vec{S}_A}{dt'}   &=  \pb{\vec{S}_A, H_{1.5\text{PN}}}  
=  \frac{2 G \sigma_A}{c^2 R^3}  \vec{L} \times \vec{S}_A    \\
&  \equiv \vec{\Omega}_A \times  \vec{S}_A,
\end{align}
from which we can conclude that the angular velocity
imparted to the spins due to the 1.5PN part of the Hamiltonian 
scales as $\vec{\Omega}_A \sim 1/c^2$. The same is not true for
$\vec{R}$ since the corresponding angular velocity due to
$ H_{1.5\text{PN}}$
(contribution to the mean motion $n$ due to $ H_{1.5\text{PN}}$
in Eq.~\eqref{mean_motion_1.5PN})
 scales as $\sim 1/c^3$.
It is due to this difference in the  nature of evolution of 
$\vec{R}$ and $\vec{S}_A$ that we should expect to 
obtain $r \sim 2$ for the former
and $r \sim 1.5$ for the latter when the procedure surrounding
Eq.~\eqref{testing} is applied to the two vectors. 
Recall that by definition, $r$ tells us how does the angular
velocity of the error vector ($\vec{R}_a - \vec{R}_n$ or 
$\vec{S_A}_a - \vec{S_A}_n$, with the subscripts
``$a$'' and ``$n$'' denoting the analytical
 and the numerical solutions, respectively) scales with $1/c$.
Our findings did match the above expectations for $\vec{R}$ and the 
spin vectors.
We did not work with $\vec{P}$ when it came to applying this
linear-fitting procedure.

\section{Summary}      
\label{summary}

With this paper, we conclude the old problem of solving the dynamics of the 1.5PN
BBH system, whose Hamiltonian was introduced in Ref.~\cite{Barker:1966zz}.
This is done for 
arbitrary masses, spins and eccentricity, and without any averaging. 
First we re-presented the solution of Ref.~\cite{Cho:2019brd} but for completeness, 
unlike Ref.~\cite{Cho:2019brd}, we included the 1PN effects too. Then we show  
how to construct the alternative AA-based 
solution in an algorithmic style. The construction of the 
latter solution requires former solution as one of the inputs.
 But the AA-based solution is superior in the sense that it can be 
extended  to 2PN order via canonical perturbation theory,
 with relative ease.
We finally introduce \texttt{BBHpnToolkit}, 
a \textsc{Mathematica} package which implements our
solutions for practical use.
We finally make comparison of the two 
analytical solutions against the 
numerical one employing linear regression, and 
the analytical solutions indeed appear to be 1.5PN accurate.

The theoretical ideas that form the foundations of the two analytical
solutions  have been presented in Refs.~\cite{Cho:2019brd, Tanay:2020gfb, Tanay:2021bff}.
Here we have assembled them, filled in the 
gaps (missing 1PN effects in Ref.~\cite{Cho:2019brd}), implemented the solutions 
in a public \textsc{Mathematica} package, thereby establishing a platform 
(both in theoretical
and practical-implementation sense) from where we can aim
to push these solutions to 2PN. Moreover, the agreement between the numerical and AA-based solution serves as a non-trivial check of all the five actions constructed in Refs.~\cite{Tanay:2020gfb, Tanay:2021bff}.

The next natural line of action would be to extend our AA-based solution to 2PN order using
canonical perturbation theory. On a parallel track, one can try to construct GWs for these
systems now that their dynamics have been solved. Adding the 2.5PN dissipative effects due to
radiation reaction to our analytic solution will also be an interesting endeavor for the future.

\acknowledgements

S.T. thanks Nicol\'as Yunes for an invitation to conduct a lecture workshop
at the University of Illinois Urbana-Champaign. The workshop served as the initial
impetus for this project. 
We also thank Nathan Johnson-McDaniel for providing useful suggestions,
Tom Colin,  Jos\'e T.~G\'alvez Ghersi,
and Laura Bernard for a careful reading of this manuscript. 
S.T. was partially supported by PSL postdoctoral fellowship.
The work of L.C.S. was partially supported by NSF
\hyphenation{CAREER}%
CAREER Award PHY-2047382 and a Sloan Foundation Research
Fellowship. R.S. acknowledges support from DST  India (Grant
No.IFA19-PH231).

\appendix

\section{Solution for flow under commuting constants}      \label{commuting_const_flow}

\subsection{Solution for flow under $H$}

The solution of flow under $H$ has been constructed in Sec.~\ref{hmsol}. 
It is mainly contained in Eqs.~\eqref{coskp1sol}, \eqref{phi_L_sol}, \eqref{S1_angle}, 
\eqref{S_2_sol}, \eqref{phi_sol}, and \eqref{p_azimuth}.

\subsection{Solution for flow under $\Eff$}

The solution of flow under $\Eff$ has been constructed in Ref.~\cite{Tanay:2021bff};
it is mainly contained in Eqs.~(A39), (A66), (A76) and (A102) of that article.
These four equations seem to determine only $\vec{R}, \vec{L}$, and $\vec{S}_1$, and 
not $\vec{S}_2$ and $\vec{P}$. But once 
we have $\vec{R}, \vec{L}$, and $\vec{S}_1$, 
determining $\vec{S}_2$ and $\vec{P}$
is quite easy. 
The former is found via $\vec{S}_2 = \vec{J} - \vec{L} - \vec{S}_2$,
 and the latter
is determined by making the following observations
 (i) $P$ is a constant under the $\Eff$ flow. So all
 we need to care about is the orientation of $\vec{P}$,
 which is dealt with in the next two bullet points.
 (ii) $\vec{L} = \vec{R} \times \vec{P}$; hence
 $\vec{L}$ is perpendicular to $\vec{R}$ and $\vec{P}$.
 (iii) under the $\Eff$ flow, the azimuthal angle 
 of $\vec{P}$ (around $\vec{L}$)
 changes by the same amount as that of $\vec{R}$.
 Bullets (ii) and (iii) let us determine the 
 orientation of $\vec{P}$ using the orientation of $\vec{R}$.
Note that the definitions and conventions of Ref.~\cite{Tanay:2021bff} don't
totally align with those followed in this paper; so care has to be taken 
when merging the results of the two papers.

\subsection{Solution for flow under $J$ }

As arrived at in Sec.~VI-B of Ref.~\cite{Tanay:2021bff}, the effect of
a flow under $J$ by an amount $\Delta \lambda$ is
increasing the azimuthal angles of $\vec{R}, \vec{P}, \vec{S}_1,$ and $\vec{S}_2$
in the IF by an amount $\Delta \lambda$.

\subsection{Solution for flow under $L$}

As also arrived at in Sec.~VI-B of Ref.~\cite{Tanay:2021bff}, the effect of
a flow under $L$ by an amount $\Delta \lambda$ is
increasing the azimuthal angles of $\vec{R}$, and $\vec{P}$
in the NIF by an amount $\Delta \lambda$.

\subsection{Solution for flow under $J_z$}

As seen in Sec.~VI-B of Ref.~\cite{Tanay:2021bff}, the effect of
a flow under $J_z$ by an amount $\Delta \lambda$ is
increasing the azimuthal angles of $\vec{R}, \vec{P}, \vec{S}_1,$ and $\vec{S}_2$
 by an amount $\Delta \lambda$, around the $z$-axis of any 
 inertial frame. IF is a special inertial frame whose $z$-axis coincides with
 $\vec{J}$.

\section{The \texttt{BBHpnToolkit} package}     \label{MMA_accuracy}

Here we give the details on some coded expressions in our \texttt{BBHpnToolkit} package
containing unnecessary higher PN order terms. Doing so does not increase the actual PN accuracy 
of our solutions since we never make use of the 2PN terms of the Hamiltonian. However doing so
may bring the analytical solutions closer to the numerical solution (obtained by numerically integrating 
the 1.5PN EOMs)
\begin{itemize}
\item As mentioned already at the end of Sec.~\ref{r-vector-sol},
 we have kept more than necessary PN accuracy in Eq.~\eqref{phi_dot}
and by extension, Eq.~\eqref{phi_sol}. All the
terms containing $\epsilon^q$ with $q \geq 2$ in Eq.~\eqref{phi_dot} are unnecessary.
\item  The derivation of the  solution Eqs.~\eqref{coskp1sol}-\eqref{jacobi-2}
 of the mutual angles between the angular momenta
for a flow under the Hamiltonian  depends on the integration 
$\int dt/r^3$; see Eq.~(3.6) of Ref.~\cite{cho2019brd}.  Here
Newtonian $r$ is enough for the desired level of accuracy, and
Eqs.~\eqref{coskp1sol}-\eqref{jacobi-2} have been derived using
Newtonian $r$ in $\int dt/r^3$. But in our 
\texttt{BBHpnToolkit}, we have coded the results obtained
by integrating 1.5PN accurate $1/r^3$. Eqs.~\eqref{r_newt} give the 1.5PN-accurate $r$. 
\end{itemize}

\section{Determining the sign of $\vec{p} \cdot \hat{r} $}     \label{sign_of_pdn}
 
  Eq.~\eqref{pdn} reads 
\begin{widetext}
\begin{align}
\vec{p} \cdot \hat{r} & = \pm \left( (2h - \epsilon(3\nu - 1)h^2) + \frac{2(1 + \epsilon(4 - \nu)h)}{r} + \frac{(-l^2 + \epsilon(6 + \nu))}{r^2} + \frac{-\epsilon \nu (l^2 + 2s_\text{eff} \cdot l / \nu)}{r^3} \right)^{1/2}   
\equiv    \pm   \sqrt{Q(r)}   .       \label{111}
\end{align}
\end{widetext}
$r^3 Q(r)$ and $Q(r)$ each have three roots:
 $r_0, r_1$, and $r_2$
(in ascending order),
where $r_0 \rightarrow 0$ in the PN limit $\epsilon \rightarrow 0$.
Since $dr/dt' = \pb{r, H^{1.5\text{PN}}} = (\vec{p} \cdot \hat{r}) Q_2$\footnote{Recall that $t'$ and $t$ represent the physical and scaled times,
respectively. The dots represent derivatives taken with respect
to the latter.}, where
$Q_2 \neq 0$ and $H^{1.5\text{PN}}$ is the 1.5PN accurate Hamiltonian,
the two roots of $\dot{r}$ (as a function of $r$) must coincide 
with two of the roots 
$r_1$ and $r_2$
of $Q(r)$.

From the quasi-Keplerian parametric (QKP) solution of 
$r$ in Sec.~\ref{r-vector-sol},
 the two roots
$r_{-}$ and $r_{+}$ of $\dot{r}$
are $a_r (1 \mp e_r)$. 
However, in general, we find that $r_{1}$ and $r_{2}$
lie outside the interval $\left[ r_{-}, r_+ \right]$. 
This non-coincidence of the roots
($(r_1, r_2) \neq (r_{-}, r_{+})$)
 can be attributed to the 
perturbative, and approximate nature of our PN calculations.
However, this leads to a certain problem of 
discontinuity when we try to determine $\vec{p} \cdot \hat{r}(t)$,
and is detailed 
further below. 
So, we need to force the coincidence of the roots
$(r_1, r_2)  = (r_{-}, r_{+})$ by hand 
to overcome this problem.

To do so, we  propose the use of 
the QKP solution for $r$ 
(in Sec.~\ref{r-vector-sol})
with modified $a_r$, and $e_r$ parameters,
in Eq.~(\ref{111}) above.
The modified parameters
 $(\tilde{a}_r, \tilde{e}_r)$ are defined as 
$\tilde{a}_r = (r_1 + r_2)/2$ and $\tilde{e}_r = (r_2  -r_1)/(r_1+r_2)$.
With this, the roots of $\dot{r}$ derived from the
correspondingly modified QKP solution,
which happen to be 
$\tilde{a}_r (1 \mp \tilde{e}_r)$,
 coincide with $r_1$, and $r_2$.

The QKP solution of $r$ implies 
$\dot{r} = a_r e_r \sin u \dot{u}$,
and $n = \dot{u} (1- e_t \cos u)$, which further implies that
for an eccentric orbit, the $t$-epochs where $\dot{r}$ vanishes are 
separated by the interval $\Delta t = \pi/n$. 
At this point, we denote by $T_0$, the unique
$t$-epoch such that $\dot{r}(t=T_0) = 0$, and  
$0 \leq T_0 < \pi/n$.
We now determine $T_0$ for it lets us
determine the correct sign in Eq.~\eqref{111} above.
Given the initial values of the phase-space variables 
$(\vec{R}, \vec{P}, \vec{S}_1, \vec{S}_2)$, 
it is a simple matter to assign
 $u^{\star} \equiv u(t=0)$ such that 
 $- \pi \leq u^{\star} \leq \pi$.
Plugging these initial values 
 in $n (t-t_0) = u- e_t \sin u$ gives us 
$t_0$. If $t_0 \geq 0$, then $T_0 = t_0$, otherwise
$T_0 = t_0 + \pi/n$.

Now finally, to determine the sign of
 $\vec{p}\cdot\hat{r}$ at a time $t_f$, we follow the following 
 easily justifiable algorithm:
\begin{enumerate}
\item Calculate $N_{P/2}$, the number of complete half-periods between $T_0$ and $t_f$. Note that one half period in $t$ is $\pi/n$.
\item Evaluate $\vec{p}\cdot\hat{r}(t=0)$ with the initial conditions.
If
    \begin{itemize}
    \item $\vec{p}\cdot\hat{r}(t=0) \leq 0$ and $N_{P/2}$ is even, or
    \item $\vec{p}\cdot\hat{r}(t=0) \geq 0$ and $N_{P/2}$ is odd,
    \end{itemize}
    then $\vec{p}\cdot\hat{r}(t_f) > 0$, and we set
    $\vec{p}\cdot\hat{r}(t_f)  =  \sqrt{Q(r)}  $.
Otherwise, $\vec{p}\cdot\hat{r}(t_f) < 0$, and we set
$\vec{p}\cdot\hat{r}(t_f)  =  - \sqrt{Q(r)}   $.
\end{enumerate}

Now we summarize the above discussion with the help of the
figure below. The broken-red curve displays the numerical solution
for $\vec{p} \cdot \hat{r}$ that we want to model. The green curve
corresponds to $\sqrt{Q}$, and it does not touch the 
x-axis because of the above-discussed non-coincidence of the roots. 
The use of the modified QKP  solution
(with $\tilde{a_r}$ and $\tilde{e_r}$) in
$\sqrt{Q(r)}$ of Eq.~\eqref{111} gives us the 
blue curve which does touch the x-axis due to the forced coincidence of the
roots. Thereafter, setting $\vec{p}\cdot\hat{r}(t_f)  =  \pm | Q(r)^{1/2}| $
as per the above bullet points,
basically amounts to flipping the blue curve about the x-axis,
but only in alternate intervals of size $\pi/n$. The result is 
 the orange curve which 
is our final emulation of the numerical red curve.
Finally note that had we not used the modified QKP solution in 
Eq.~\eqref{111} above, we would have ended up with the broken 
black curve as our final model, which is 
obtained by flipping the green curve instead.
 As is evident from the figure, this broken black
 curve suffers from the problem
of discontinuity that we alluded to in the beginning.
This concludes our discussion of determining the 
sign of $\vec{p} \cdot \hat{r}$.

\begin{widetext}

\begin{figure}[h!]
    \centering
            \includegraphics[width= \textwidth]{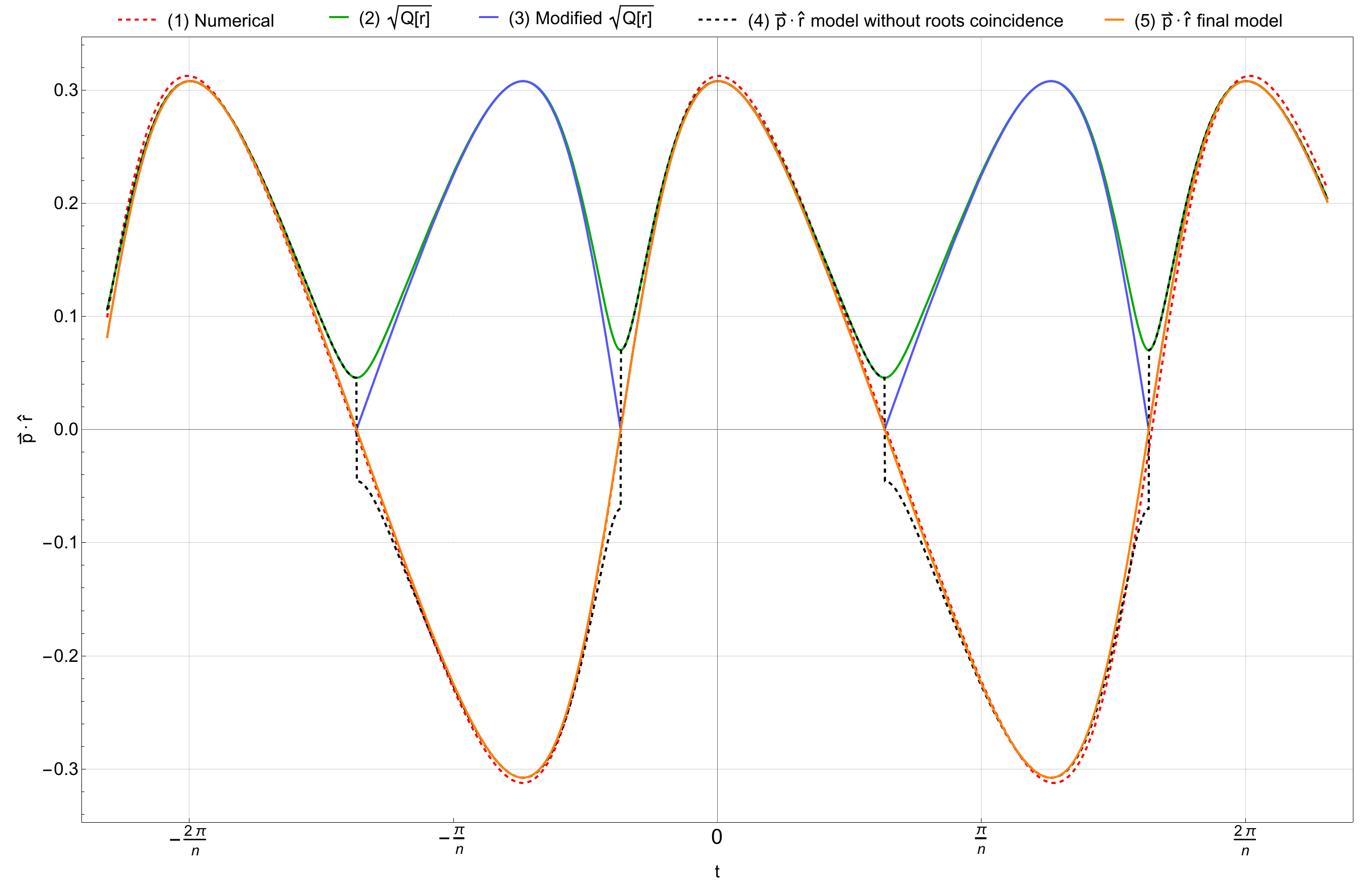} 
        \caption{Comparison of different implementations of 
        $\vec{p} \cdot \hat{r}$.  
Curve (1) shows $\vec{p} \cdot \hat{r}$ obtained via numerical integration.  
Curves (2) and (3) display $|\vec{p} \cdot \hat{r}|$ as per the standard QKP
solution, and the modified QKP solution, respectively.
Curves (4), and (5) display $\vec{p} \cdot \hat{r}$ obtained by flipping 
Curves (2) and (3) about the x-axis within alternate time intervals.
Curve (5) is the final analytical model that we propose, and also
the one which is closer to the numerical Curve (1).}
        \label{fig:image1}
\end{figure}

\end{widetext}


\bibliography{pn_v2}

\end{document}